\documentclass[preprintnumbers,amsmath,amssymb,showpacs,twocolumn]{revtex4}

\usepackage{graphicx}
\usepackage{dcolumn}
\usepackage{bm}

\begin{document}

\title{ Simultaneous quantum state teleportation via the locked entanglement channel}

\author{Meiyu Wang $^a$}
\author{Fengli Yan $^a$ }
\thanks {Corresponding author. Email address: flyan@mail.hebtu.edu.cn}
\author{ Ting Gao
$^b$}
\author{Youcheng Li $^a$}
\affiliation {$^a$ College of Physics  and Information Engineering, Hebei Normal University, Shijiazhuang
050016,
China\\
$^b$ College of Mathematics and Information Science, Hebei
  Normal University, Shijiazhuang 050016, China}

\date{\today}

\begin{abstract}
  We present a simultaneous quantum state teleportation scheme, in
  which receivers can not recover their quantum state separately. When
  they want to recover their respective quantum state, they must perform an
  unlocking operator together.
\end{abstract}


\maketitle

Quantum entanglement plays an important role in various fields of quantum information, such as quantum
computation, quantum cryptography, quantum teleportattion and dense coding etc. Quantum teleportation is one of
the most important applications of quantum entanglement. In quantum teleportation process, an unknown state can
be transmitted from a sender Alice to a receiver Bob without transmission of carrier of quantum state. Since
Bennett et al \cite {s1} presented a quantum teleportation scheme, there has been great development in
theoretical and experimental studies. Now quantum teleportation has been generalized to many cases \cite
{s2,s3,s4,s5,s6,s7,s8,s9}. Moreover, quantum teleportation has been demonstrated with the polarization photon
\cite {s10} and a single coherence mode of fields \cite {s11} in the experiments. The teleportation of a
coherent state corresponding to continuous variable system was also realized in the laboratory \cite {s12}.

In the original proposal, the teleportation of a single qubit
$|\phi\rangle=a|0\rangle+b|1\rangle$ is executed as follows: Alice
and Bob initially  share an EPR pair as a quantum channel. Alice
performs a joint measurement on the composed system (qubit to be
teleported  and one of the entangled pair). She transmits the
outcome to Bob through a classical channel. Bob applies a
corresponding unitary operation on his particle of the entangled
pair, which is chosen in accordance with the outcome of joint
measurement. The final state of Bob's qubit is completely equivalent
to the original unknown state.

If Alice wants to teleport the quantum state $|\phi_1\rangle$ to Bob and teleport $|\phi_2\rangle$ to Charlie,
obviously two EPR pairs are required. One is shared between Alice and Bob; the other is shared between Alice and
Charlie. The quantum state teleportation can be completed by Bennett's protocol. But if Bob and Charlie want to
receive their respective quantum state simultaneously, how do they complete the teleportation? In this paper, we
will present a simultaneous quantum state teleportation scheme. In this scheme, the recievers Bob and Charlie
can synchronously recover the quantum state which Alice teleported to them respectively by locking their quantum
channels.

 To present our scheme clearly, let's first begin with simultaneous
 quantum state teleportation between one sender and two receivers.

 The quantum states of two qubits $T_1$ and $T_2$ to be teleported
 are  as follows
 \begin{equation}
\begin{array}{l}|\phi_1\rangle_{T_1}=\alpha_1|0\rangle_{T_1}+\beta_1|1\rangle_{T_1},\\
|\phi_2\rangle_{T_2}=\alpha_2|0\rangle_{T_2}+\beta_2|1\rangle_{T_2},\\
 \end{array}\end{equation}
Alice wants to teleport $|\phi_1\rangle$ to Bob and $|\phi_2\rangle$
to Charlie simultaneously. Suppose that Alice, Bob and Charlie share
two EPR pairs denoted as
\begin{equation}
\begin{array}{l}|EPR\rangle_1=\frac {1}{\sqrt
2}(|00\rangle_{A_1B}+|11\rangle_{A_1B}),\\
|EPR\rangle_2=\frac {1}{\sqrt
2}(|00\rangle_{A_2C}+|11\rangle_{A_2C}),\\
\end{array}
\end{equation}
where $A_1$ and $A_2$ belong to Alice, $B$ and $C$ belong to Bob and
Charlie respectively. Then the quantum state of the joint system
(qubits to be teleported and two EPR pairs) can be written as
\begin{equation}
\begin{array}{l}
|\Phi\rangle=(\alpha_1\alpha_2|00\rangle+\alpha_1\beta_2|01\rangle+\alpha_2\beta_1|10\rangle+\beta_1\beta_2|11\rangle)_{T_1T_2}\\
~~~~~~~~\otimes\frac
{1}{2}(|0000\rangle+|0101\rangle+|1010\rangle+|1111\rangle)_{A_1A_2BC}.
\end{array}
\end{equation}

The scheme of simultaneous teleportation consists of the following
five steps.

(S1) Locking the quantum channels

In this step of teleportation, Bob and Charlie perform a joint
unitary transformation
\begin{equation}
U_{BC}=\frac {1}{\sqrt 2}\left
[\begin{array}{cccc}1&0&1&0\\
0&1&0&1\\
0&1&0&-1\\
1&0&-1&0\\\end{array}\right ]
\end{equation}
on the particles $B$ and $C$. After that, the state of the total
system becomes
\begin{equation}
\begin{array}{l}
~~~~|\Phi'\rangle\\
=(\alpha_1\alpha_2|00\rangle+\alpha_1\beta_2|01\rangle+\alpha_2\beta_1|10\rangle+\beta_1\beta_2|11\rangle)_{T_1T_2}\\
~~~\otimes\frac {1}{2\sqrt
2}(|0000\rangle+|0011\rangle+|0101\rangle+|0110\rangle\\
~~~~+|1000\rangle-|1011\rangle+|1101\rangle-|1110\rangle)_{A_1A_2BC}.
\end{array}
\end{equation}

 (S2) Performing  Bell-basis
measurement

Alice performs a projective measurement on $A_1T_1$ and one on $A_2T_2$ in the Bell-basis $\{|\Psi_m\rangle,
m=0,1,2,3\}$, where
\begin{equation}
\begin{array}{cc}
|\Psi_0\rangle=\frac {1}{\sqrt 2}(|00\rangle+|11\rangle), &
|\Psi_1\rangle=\frac {1}{\sqrt 2}(|01\rangle+|10\rangle),\\
|\Psi_2\rangle=\frac {1}{\sqrt 2}(|00\rangle-|11\rangle), &
|\Psi_3\rangle=\frac {1}{\sqrt
2}(|01\rangle-|10\rangle).\\
\end{array}\end{equation}
Let $\sigma_j$   be  one  member of the set of rotation matrices
\begin{equation}
\{I, \left (\begin{array}{cc}0&1\\
1&0\\\end{array}\right), \left (\begin{array}{cc}1&0\\
0&-1\\\end{array}\right), \left (\begin{array}{cc}0&1\\
-1&0\\\end{array}\right)\}
\end{equation}
 composed  of   the identity operator and three Pauli spin
operators. It is easy to prove that  Eq.(5) can be rewritten as
\begin{equation}
\begin{array}{l}
|\Phi'\rangle=\Sigma_{m,n}|\Psi_m\rangle_{A_1T_1}|\Psi_n\rangle_{A_2T_2}
U_{BC}\sigma_m|\phi_1\rangle_B\otimes\sigma_n|\phi_2\rangle_C.
\end{array}
\end{equation}
After the projective measurement, the state of particles $B$ and $C$
 collapses into
\begin{equation}
|\Psi\rangle_{BC}=U_{BC}\sigma_m|\phi_1\rangle_B\otimes\sigma_n|\phi_2\rangle_C,
\end{equation}
which corresponds to the measurement result $|\Psi_m\rangle,
|\Psi_n\rangle$.

(S3) Transmitting the measurement outcome

After performing the Bell-basis measurement, Alice transmits the
outcome of measurement (i.e. $m$ and $n$) to Bob and Charlie.

(S4) Unlocking the quantum state

According to Eq.(9) the quantum state of particle $B$ and $C$ is locked, so Bob and Charlie must "unlock" $B$
and $C$ firstly. In order to recover the quantum states which are teleported from Alice, Bob and Charlie
perform a unitary operator $U^+_{BC}$ on the particles $B$ and $C$, where
\begin{equation}
U^+_{BC}=\frac {1}{\sqrt 2}\left
[\begin{array}{cccc}1&0&0&1\\
0&1&1&0\\
1&0&0&-1\\
0&1&-1&0\\\end{array}\right ].
\end{equation}
After that  the state of particle $B$ and $C$ is  transformed into
\begin{equation}
\begin {array}{l}~~~|\Psi'\rangle_{BC}\\
=U^+_{BC}U_{BC}\sigma_m|\phi_1\rangle_B\otimes\sigma_n|\phi_2\rangle_C
=\sigma_m|\phi_1\rangle_B\otimes\sigma_n|\phi_2\rangle_C.
\end{array}\end{equation}

(S5) Recovering  the quantum state

 Bob and Charlie perform a local unitary operator $\sigma_m$ and
$\sigma_n$ on particles $B$ and $C$ respectively, then  they  will
obtain $|\phi_1\rangle$ and $|\phi_2\rangle$ immediately.

In the following, we will discuss why we call the step (S1) "locking" the quantum channel. It is easy to verify
that $U_{BC}$ can be realized by a Hadamard operator on particle $B$ and a CNOT operator ($B$ is control qubit
and $C$ is target qubit). Apparently $U_{BC}$ is a non-local operator. Now, is there any  entanglement between
$B$ and $C$ after Bob and Charlie make  the unitary operator $U_{BC}$ on them? The answer is negative. To
investigate the entanglement between $B$ and $C$, we employ the Peres-Horodecki criterion \cite {s13,s14} for
two qubits that their density operator $\rho$ is separable if and only if its partial transposition is positive.
The partial transposition of $\rho$ is defined as
\begin{equation}
\rho^{T}=\Sigma_{ijkl}\rho_{jikl}|i\rangle\langle j|\otimes
|k\rangle\langle l|,
\end{equation}
where $\rho=\Sigma_{ijkl}\rho_{ijkl}|i\rangle\langle j|\otimes
|k\rangle\langle l|.$ After transformation, the reduced density
operator of $(B,C)$ is given as
\begin{equation}
\rho'_{BC}=\frac {1}{4}I.
\end{equation}
The partial transposition of $\rho'_{BC}$ has only positive
eigenvalues $\frac {1}{4},\frac {1}{4},\frac {1}{4},\frac {1}{4}$.
The result implies that there is no entanglement between $B$ and
$C$. In fact, before transformation the reduced density operator of
$(B,C)$ is $\rho_{BC}=\frac {1}{4}I$. For any unitary operator $U$
we always have
\begin{equation}
\rho'_{BC}=U^+\rho_{BC}U=\rho_{BC}=\frac {1}{4}I,
\end{equation}
i.e. there is  no entanglement between $B$ and $C$ for arbitrary
unitary operator $U$. Notice that after transformation, the quantum
channels become
\begin{equation}
\begin{array}{l}
|\Psi'\rangle=\frac {1}{2\sqrt
2}(|0000\rangle+|0011\rangle+|0101\rangle+|0110\rangle\\
~~~~~~~~~~~~+|1000\rangle-|1011\rangle+|1101\rangle-|1110\rangle)_{A_1A_2BC}\\
~~~~~=\frac {1}{2\sqrt
2}[(|00\rangle+|10\rangle)_{A_1B}\otimes(|00\rangle+|11\rangle_{A_2C})\\~~~~~~~~~~~~+(|01\rangle-|11\rangle)_{A_1B}
\otimes(|01\rangle+|10\rangle_{A_2C})].\\
\end{array}
\end{equation}
Eq. (15) shows that  $|\Psi'\rangle$ is maximally entangled state of $A_1B$ and $A_2C$. In other words, the
function of $U_{BC}$ is not to entangle $(B,C)$ but $(A_1B, A_2C)$. In some sense, $U_{BC}$ is like a "lock"
which prevents Bob and Charlie from recovering their quantum states separately. More surprisingly, the initial
state of $(A_1, B)$ is maximally entangled, while after  performing the unitary transformation $U_{BC}$ the
reduced density operator $\rho'_{A_1B}$ reads
\begin{equation}
\rho'_{A_1B}=\frac {1}{4}\left
[\begin{array}{cccc}1&0&1&0\\
0&1&0&-1\\
1&0&1&0\\
0&-1&0&1\\\end{array}\right ].
\end{equation}
The partial transposition of $\rho'_{A_1B}$ has only nonnegative eigenvalues $0,0,\frac {1}{2}, \frac {1}{2}$.
The result implies that the entanglement of $(A_1, B)$ has vanished. Similarly, the entanglement of $(A_2, C)$
has also vanished. That is, $A_1B$ and $A_2C$ have no function of the quantum channels, after $U_{BC}$ is
applied on $B$ and $C$ particles. Alice can teleport $|\phi_1\rangle$ and $|\phi_2\rangle$ to Bob and Charlie
respectively due to the entanglement between $A_1B$ and $A_2C$.

It is worthy  pointing out that "locking" the quantum channels can be completed by Alice. On the one hand,
before distributing the entanglement pairs, Alice does the transformation $U_{BC}$ on the two qubits which will
be sent to Bob and Charlie, i.e. on qubits B and C. The advantage of which is that Bob and Charlie need not come
together to lock the states. They only need to come together at the time they want to unlock it. On the other
hand, Alice may also make unitary operator $U_{A_1A_2}$ on qubits $A_1$ and $A_2$ after distributing the
entanglement pairs
\begin{equation}
\begin{array}{l}~~U_{A_1A_2}\otimes I_{BC}\frac
{1}{2}(|0000\rangle+|0101\rangle+|1010\rangle+|1111\rangle)_{A_1A_2BC}\\
=\frac {1}{2\sqrt 2}
(|0000\rangle+|0011\rangle+|0101\rangle+|0110\rangle+|1000\rangle\\
~~~~-|1011\rangle+|1101\rangle-|1110\rangle)_{A_1A_2BC},\\
\end{array}
\end{equation}
where
\begin{equation}
U_{A_1A_2}=\frac {1}{\sqrt 2}\left
[\begin{array}{cccc}1&0&0&1\\
0&1&1&0\\
1&0&0&-1\\
0&1&-1&0\\\end{array}\right ].
\end{equation}

 It is straightforward to generalize the above protocol to $n$
 receivers who recover the quantum states simultaneously. Suppose
 Alice wants to teleport $|\phi_i\rangle$ to Bob $i$
$(i=1,2,\ldots, n),$ Alice and every receiver share an EPR pair
denoted as
\begin{equation}
|EPR\rangle_i=\frac {1}{\sqrt
2}(|00\rangle_{A_iB_i}+|11\rangle_{A_iB_i}), i=1,2,\ldots,n.
\end{equation}
The steps in this case are similar to that of two receivers. It is
only necessary to replace locking operator $U_{B_1B_2}$ and
unlocking operator $U^+_{B_1B_2}$ by unitary operator
\begin{equation}
U_{B_1B_2\cdots B_n}=\Pi^n_{i=2}({\textrm{CNOT}})_{B_1B_i}H_{B_1}
\end{equation}
and $U^+_{B_1B_2\cdots B_n}$ respectively. Here $H_{B_1}$ is a
Hadmard operator on the qubit $B_1$.

Without difficulty one can figure  out the network for locking the
$n$ quantum channels, for saving  space we do not depict it here.

In summary, we present a simultaneous quantum state teleportation scheme, in which receivers can not recover
their quantum state separately. When they want to recover their quantum states, they must perform a unlocking
operator together.

 \acknowledgments  This work was supported by the National  Natural Science Foundation of
China under Grant No: 10671054,  Hebei Natural Science Foundation of China under Grant Nos: A2005000140, 07M006
and the  Key Project of Science and Technology Research of Education Ministry of China under Grant No:207011.

\end{document}